\documentclass[12 pt]{article}

\usepackage{amsmath,amssymb,amsfonts,latexsym,float,graphics,epsfig}

\usepackage{setspace}
\usepackage[top=2cm, bottom=2cm, left=2cm, right=2cm]{geometry}
\usepackage[colorlinks]{hyperref}
\begin{document}

\renewcommand\theequation{\arabic{section}.\arabic{equation}}
\catcode`@=11 \@addtoreset{equation}{section}\newtheorem{axiom}{Definition}[section]
\newtheorem{theorem}{Theorem}[section]
\newtheorem{axiom2}{Example}[section]
\newtheorem{lem}{Lemma}[section]
\newtheorem{prop}{Proposition}[section]
\newtheorem{cor}{Corollary}[section]
\newtheorem{remark}{Remark}
\newcommand{\be}{\begin{equation}}
\newcommand{\ee}{\end{equation}}

\newcommand{\equal}{\!\!\!&=&\!\!\!}
\newcommand{\rd}{\partial}
\newcommand{\g}{\hat {\cal G}}
\newcommand{\bo}{\bigodot}
\newcommand{\res}{\mathop{\mbox{\rm res}}}
\newcommand{\diag}{\mathop{\mbox{\rm diag}}}
\newcommand{\Tr}{\mathop{\mbox{\rm Tr}}}
\newcommand{\const}{\mbox{\rm const.}\;}
\newcommand{\cA}{{\cal A}}
\newcommand{\bA}{{\bf A}}
\newcommand{\Abar}{{\bar{A}}}
\newcommand{\cAbar}{{\bar{\cA}}}
\newcommand{\bAbar}{{\bar{\bA}}}
\newcommand{\cB}{{\cal B}}
\newcommand{\bB}{{\bf B}}
\newcommand{\Bbar}{{\bar{B}}}
\newcommand{\cBbar}{{\bar{\cB}}}
\newcommand{\bBbar}{{\bar{\bB}}}
\newcommand{\bC}{{\bf C}}
\newcommand{\cbar}{{\bar{c}}}
\newcommand{\Cbar}{{\bar{C}}}
\newcommand{\Hbar}{{\bar{H}}}
\newcommand{\cL}{{\cal L}}
\newcommand{\bL}{{\bf L}}
\newcommand{\Lbar}{{\bar{L}}}
\newcommand{\cLbar}{{\bar{\cL}}}
\newcommand{\bLbar}{{\bar{\bL}}}
\newcommand{\cM}{{\cal M}}
\newcommand{\bM}{{\bf M}}
\newcommand{\Mbar}{{\bar{M}}}
\newcommand{\cMbar}{{\bar{\cM}}}
\newcommand{\bMbar}{{\bar{\bM}}}
\newcommand{\cP}{{\cal P}}
\newcommand{\cQ}{{\cal Q}}
\newcommand{\bU}{{\bf U}}
\newcommand{\bR}{{\bf R}}
\newcommand{\cW}{{\cal W}}
\newcommand{\bW}{{\bf W}}
\newcommand{\bZ}{{\bf Z}}
\newcommand{\Wbar}{{\bar{W}}}
\newcommand{\Xbar}{{\bar{X}}}
\newcommand{\cWbar}{{\bar{\cW}}}
\newcommand{\bWbar}{{\bar{\bW}}}
\newcommand{\abar}{{\bar{a}}}
\newcommand{\nbar}{{\bar{n}}}
\newcommand{\pbar}{{\bar{p}}}
\newcommand{\tbar}{{\bar{t}}}
\newcommand{\ubar}{{\bar{u}}}
\newcommand{\utilde}{\tilde{u}}
\newcommand{\vbar}{{\bar{v}}}
\newcommand{\wbar}{{\bar{w}}}
\newcommand{\phibar}{{\bar{\phi}}}
\newcommand{\Psibar}{{\bar{\Psi}}}
\newcommand{\bLambda}{{\bf \Lambda}}
\newcommand{\bDelta}{{\bf \Delta}}
\newcommand{\p}{\partial}
\newcommand{\om}{{\Omega \cal G}}
\newcommand{\ID}{{\mathbb{D}}}
\newcommand{\pr}{{\prime}}
\newcommand{\prr}{{\prime\prime}}
\newcommand{\prrr}{{\prime\prime\prime}}

\title{On Geometrical Couplings of Dissipation and Curl Forces} \maketitle
\begin{center}

O\u{g}ul Esen\footnote{E-mail: oesen@gtu.edu.tr}\\
Department of Mathematics, \\ Gebze Technical University, 41400 Gebze,
Kocaeli, Turkey.

\bigskip

Partha Guha\footnote{E-mail: partha.guha@ku.ac.ae}\\
Department of Mathematics,
Khalifa University\\
P.O. Box 127788, Zone -1 Abu Dhabi, UAE. 

\bigskip

Hasan G\"{u}mral\footnote{E-mail: hgumral@yeditepe.edu.tr}\\
Department of Mathematics, \\ 
Yeditepe University, 34755 Ata{\c{s}}ehir, {\.{I}}stanbul, Turkey.

\end{center}

\date{ }

\smallskip

\smallskip

\begin{center}

{\it Dedicated to Sir Michael Berry on his 80th birthday with great respect and admiration.}

\end{center}

\bigskip

\begin{abstract}
\textit{In this paper, we present several geometric ways to incorporate gyroscopic and dissipative forces to curl forces. We first present a proper metriplectic geometry. Then, using the Herglotz principle
and generalized Euler-Lagrange equation, we propose a formulation of dissipative radial curl forces. Finally, we extend our result to azimuthal curl force using Galley's method,
which leads to a natural formulation for Lagrangian and Hamiltonian dynamics of generic non-conservative systems.
\paragraph{MSC classes:} 34D20, 37C20, 37C60
}
\end{abstract}

\tableofcontents
\setlength{\parskip}{4mm}

\onehalfspacing

\section{Introduction}

In a series of papers \cite{BS466,BS2,BS3}, Berry and Shukla have introduced Newtonian dynamics driven by forces, depending only on position,  whose curl is non-zero. Assuming unit mass for convenience, we record the dynamics generated by so called a curl force as
\begin{equation} \label{curl-f}
\ddot{\bf r} = \textbf{F}({\bf r}), \qquad \nabla \times \textbf{F}({\bf r}) \neq 0.
\end{equation}
Note that, the motion governed by a curl force is non-conservative, that is the work done by $\textbf{F}$ depends on the path. But, apart from velocity-dependent frictional forces, curl forces are not dissipative. If there exists no attractors, it is immediate to realize that, a flow generated by a curl force preserve the volume in the position-velocity phase space $({\bf r}, {\bf v})$. The theory of curl forces has found profound applications in optics, laser physics
and anisotropic Kepler problem \cite{BS2,CNV,AML,SS,Gutz,Devaney}. 

Independently, the concept of curl force has appeared in some works \cite{Lundmark1,Lundmark2,Lundmark3} of  Link\"oping School  by Rauch-Woiciechowski, Marciniak and Lundmark.
Practically, Link\"oping School studied the most general version of the construction of curl forces. 

Curl forces neither can be  derived from a scalar potential nor, in general, can be examined in the realm of Hamiltonian
Mechanics.
Nevertheless, a large class of such non-conservative forces (though not all) can be generated from Hamiltonians of a special type, in which
kinetic energy is an anisotropic quadratic function of momentum. Berry and Shukla \cite{BS3} listed various permitted classes of 
curl forces which are kinetically quadratic and anisotropic. Our investigation in this work starts precisely at this point. 

In the present work, our goal is to extend linear (Hamiltonian) curl force formalism to a 
generalized theory by incorporating dissipative and gyroscopic forces. In order to couple such dissipative terms to the system \eqref{curl-f}, we employ pure geometric/algebraic techniques  available in the literature, such as double bracket dissipations, contact Hamiltonian formalism and Galley's approach. To fix the notation and for the sake of the completeness of this work, in Section \ref{can-Ham}, we shall recall some basics on Hamiltonian formulations of curl forces. It is Section \ref{met-Ham} where the first trial through the coupling of dissipative terms to curl forces will be presented by means of a proper metriplectic bracket. To have this geometrically, we shall realize the canonical Hamiltonian formalism as a coadjoint flow then apply the double bracket dissipation approach to this realization. In this section, we shall also address symplectic two-forms involving magnetic terms. This leads to couple gyroscopic forces to the curl force model. In Section \ref{Contact}, we shall provide contact Hamiltonian realization of the curl force model and, particularly, study
dissipative curl forces using the generalized Euler-Lagrange equations formulated by Gustav Herglotz. This will lead us to analyse the 
variational aspects, as well as Lagrangian formulation, of the dissipative terms. An alternative method to employ dissipative Lagrangian dynamics is Galley's approach.
In Section \ref{Galley-Sec}, after briefly summarizing Galley's approach for dissipative Lagrangian-Hamiltonian dynamics. We shall incorporate this approach to the case of curl force models.  We conclude this paper with a modest outlook.

\section{Hamiltonian Realization of Curl Forces}\label{can-Ham}

\subsection{Hamiltonian Dynamics}
Let $\mathcal{P}$ be a Poisson manifold equipped with a Poisson bracket $\{\bullet , \bullet \}$, \cite{LaPi12,Va12} a skew-symmetric algebra on the space of real valued functions satisfying the Jacobi and the Leibniz identities. 
For a given Hamiltonian function $H$, the dynamcs of an observable $F$ are determined through Hamilton's equations
\begin{equation}\label{HamEq}
\dot{F}=\{F,H\}=X_{H}(F),
\end{equation}
where $X_{H}$ is the Hamiltonian vector field. 
Skew-symmetry of Poisson bracket reads that Hamiltonian function is a conserved quantity. In some classical models, Hamiltonian function is considered to be the total mechanical energy, so that the equality $\dot{H}=0$ states the conservation of energy. This manifests  the reversible character of Hamiltonian dynamics. 

Alternatively, a Poisson structure can be introduced by means of a bivector field. To have this formulation, start with a Poisson bracket, and define a bivector field $\Lambda$ according to the  identity
\begin{equation} \label{bivec-PoissonBra}
\Lambda(dF,dH):=\{F,H\}
\end{equation}
for all real valued functions $F$ and $H$ on $\mathcal{P}$. Here, $dF$ stands for the exterior (de-Rham) derivative of $F$. Accordingly, a Poisson manifold can be defined by a tuple $(\mathcal{P},\Lambda)$. In this case, the Jacobi identity is manifested in terms of the Schouten-Nijenhuis bracket \cite{deAzPeBu96} that is,
\begin{equation} \label{Poisson-cond}
[\Lambda,\Lambda]=0.
\end{equation} 
Inversely, for a bivector field commuting with itself under the Schouten-Nijenhuis algebra, one can define a bracket of functions $\{\bullet,\bullet\}$ satisfying the requirements of being a Poisson bracket. This is simply achieved by referring to the identity \eqref{bivec-PoissonBra} but from the reverse order.

Consider  a Poisson manifold $\mathcal{P}$. 
The characteristic distribution, that is the image space of all Hamiltonian vector fields is integrable. This reads (symplectic) foliation of $\mathcal{P}$, \cite{wei83}. In this foliation, on each leaf, the Poisson bracket turns out to be non-degenerate and determines a symplectic two-form. This latter argument manifests that every symplectic manifold is a Poisson manifold. In the present work, we are interested in the following particular instance. Assume a $2$-dimensional manifold  $\mathcal{Q}$ equipped with a coordinate chart $\textbf{r}=(x,y)$. Being cotangent bundle, $\mathcal{P}=T^*\mathcal{Q}$ is a symplectic manifold hence a $4$-dimensional Poisson manifold \cite{Gumral}. Here, by taking the momenta $\textbf{p}=(p_x,p_y)$, the induced Poisson bracket, for two functions $F$ and $H$, is computed to be 
\begin{equation}
\{F,H\}(\textbf{q},\textbf{p}) = \frac{\partial F}{\partial x}\frac{\partial H}{\partial p_x} + \frac{\partial F}{\partial y}\frac{\partial H}{\partial p_y} - \frac{\partial F}{\partial p_x} \frac{\partial H}{\partial x} - \frac{\partial F}{\partial p_y} \frac{\partial H}{\partial y}.
\end{equation}
Recalling from \cite{BS466}, in the present work, we are interested in Hamiltonian functions in the form
\begin{equation}\label{Ham-F}
 H({\bf q}, {\bf p}) = \frac{1}{2}(p_{x}^{2} - p_{y}^{2}) + U({\bf q}).
 \end{equation}
The Hamiltonian function \eqref{Ham-F} generates a Newtonian dynamics induced by a curl force. Before examine this, we encounter two different types of curl forces according to the functional structure of the potential $U({\bf q})$.  Then a direct calculation proves that the force field is curl free if the mixed second partial derivative of the potential function is non vanishing. 

\subsection{Radial Curl Forces}
We first consider the Hamiltonian function $H$ in \eqref{Ham-F} where the potential energy $U$ is a function of a saddle with equal principal curvatures \cite{Guha1}. That is, we consider $U$ as a function of $\xi=(1/2)(x^2-y^2)$. Let us record this particular instance for future reference as follows
\begin{equation}\label{E1a}
H=\frac{1}{2}(p_x^2-p_y^2) +U\big(\frac{1}{2}(x^2-y^2)\big).
\end{equation}
The Hamilton's equation can be computed through \eqref{HamEq}. Accordingly, by properly coupling the differential equations one can easily obtain Newton's equations as
\begin{equation} 
\begin{split}
\ddot{x} &=\dot{p}_x=-xU^\prime\big(\frac{1}{2}(x^2-y^2)\big)=F_x, \\
\ddot{y} &=-\dot{p}_y=-yU^\prime\big(\frac{1}{2}(x^2-y^2)\big)=F_y . 
\end{split}
\end{equation}
In this case, the curl of the generating force field $\textbf{F}=(F_x,F_y)$ is computed to be
\begin{equation} 
\nabla \times \textbf{F}=-2xyU^{\prime\prime}\big(\frac{1}{2}(x^2-y^2)\big)\textbf{k}
\end{equation}
and, evidently, it is non-vanishing as long as the second derivative $U^{\prime\prime}(\xi)$ is not zero. To explore the nature of the curl force, we employ the polar coordinates $x=r\cos\theta$ and $y=r\sin\theta$ with corresponding unit directions given by $(\textbf{r}, \boldsymbol{\phi} )$.
In this realization, the force field has the following appearance 
\begin{equation}
\textbf{F}=-rU^\prime(\frac{1}{2}r^2\cos2\phi)\textbf{r}
\end{equation}
showing that it is radially directed. This formulation of the force field manifests that the angular momentum is a constant of motion. The conservation of the angular momentum reads that there is no torque about the
origin for the radial curl forces.  So that, there are two conserved quantities, the energy and the angular momentum.   
Hence, we can argue that the radial curl forces are integrable. 

It is known for almost a century  that the equilibrium position of the particle
becomes stable if the surface rotates around the vertical axis sufficiently fast. This can be demonstrated 
in the following sense. Consider the saddle force field $ \big(F_x = x, F_y = -y\big)$,
 make it time-dependent by rotating each vector counterclockwise with an angular velocity, which
describes the motion of the unit point mass in this force field \cite{Guha2}.
Earnshaw’s theorem states that trapping of a charged particle cannot be 
achieved with a static potential,
because any static potential, which fulfills Laplace’s law, lacks potential minimal. Microscopic particle 
traps usually use time dependence to work round Earnshaw's theorem. The trapping of the particle can 
be achieved by rotating the saddle potential around the axis with a suitable angular frequency, this was 
idea of Wolfgang Paul \cite{Paul}. Paul's idea was to stabilize the
saddle by ``vibrating'' the electrostatic field, by analogy with
the so called Stephenson-Kapitsa \cite{Stephenson,KapitsaPen} pendulum. Nobel Prize in physics was awarded to W. Paul 
for his invention of the trap. A mathematically formal analysis of
stable trapping by using the analysis of the Mathieu equation allows to formulate the limitations of the 
effective pseudopotential description.  The Mathieu stability limit can be reached for a few charges in the trap.
Recently Kirillov and Levi \cite{KL1,KL2} demonstrated that rotating saddle potentials exhibit precessional motion
due to a hidden Coriolis-like force. Dynamics driven by curl forces has been studied by engineers and applied
mathematicians, it was not so popular among physicists, although it has very beautiful applications
in central force and nonlinear dynamics \cite{BS2,GCGP,Guha}.

\subsection{Azimuthal Curl Forces.} 

As a second particular case of the Hamiltonian function \eqref{Ham-F}, we take that the potential is a function of $\xi=xy$. In this case, we write the Hamiltonian function as 
\begin{equation}\label{E1b}
H=\frac{1}{2}(p_x^2-p_y^2) +U(xy).
\end{equation}
In this case, Newton's equations turn out to be 
\begin{equation}
\ddot{x} = -yU^{\prime}(xy), \quad \ddot{y} = xU^{\prime}(xy).
\end{equation}
Such forces are called azimuthal curve forces. To justify this labelling it is enough to see that, in the polar coordinates, the force field admits only angular unit vector 
\begin{equation}
 {\bf F}({\bf r})
= rU^{\prime}(\frac{1}{2}r^2 \sin(2\phi))\boldsymbol{\phi}.
\end{equation}
In general, the energy, that is the Hamiltonian function \eqref{E1b}, is the only conserved quantity. So that, this class of azimuthal forces is non-integrable. 
This is connected to optical vortex curl force for $U(xy) = xy$. This case possesses the rotational symmetry.  One must note that for various
non-central forces ${\bf F}({\bf r}) = F_{\phi}(r)\boldsymbol{\phi}$ do not fall into this Hamiltonian class for
general $F_{\phi}(r)$.

\begin{remark}
There exists another type of curl forces which are directed in one direction in the Cartesian coordinates, but depending on both of
the coordinates. These are known as the shear curl forces.
The special classes of shear forces can be obtained from the Hamiltonians $H$ in form \eqref{Ham-F} where the potential is $U=U(x\pm y)$. This force fields are out of our scope in this work.
\end{remark}

\subsection{Kapitsa-Merkin Non-conservative Positional Forces} 

The curl forces picture is related to the theory of Kapitsa-Merkin non-conservative positional
forces. The linearised dynamics of a rotating shaft formulated by Kapitsa \cite{Kapitsa} is given by
\begin{equation} \label{Kapitsa}
\ddot{x} + ay + bx = 0, \qquad \ddot{y} - ax + by = 0.
\end{equation}
The corresponding characteristic equation shows that the addition of a non-zero non-conservative 
curl force ( i.e. $ a \neq 0$) to a stable system with a stable potential energy makes it unstable.
This is connected to Merkin's result \cite{Merkin,Merkin1}, which states that the introduction of non-conservative linear forces
into a system with a stable potential and with equal frequencies destroys the stability regardless of the form
of non-linear terms. It is worth mentioning that the positional force, i.e. 
the terms $ay$ and $-ax$ are proportional to $\omega^2$,
where $\omega$ is the rotation rate of the shaft. Let us write (\ref{Kapitsa}) as
\begin{equation}
\begin{pmatrix}
\ddot{x} \\
\ddot{y}
\end{pmatrix}+
\begin{pmatrix}
b& 0\\
0 & b
\end{pmatrix}
\begin{pmatrix}
x \\
y
\end{pmatrix}
+\begin{pmatrix}
0& a\\
-a & 0
\end{pmatrix}
\begin{pmatrix}
x \\
y
\end{pmatrix}
=\begin{pmatrix}
0 \\
0
\end{pmatrix}.
\end{equation}
where the potential part, say $B$, is a diagonal matrix with equal
eigenvalues $b$ whereas non-conservative part, say $A$, is an skew-
symmetric matrix. It is easy to see from the corresponding characteristic equation $\hbox{ det }\big((\lambda^2 + b){\Bbb I} + A \big) = 0$ 
that $\lambda^2 + b$ is imaginary, thus we say that $\lambda$ is unstable. 

It is possible to recast the dynamical equation in \eqref{Kapitsa} as a Hamiltonian curl force system as follows. 
Consider the following Hamiltonian function 
\begin{equation} \label{Hammam}
H = \frac{1}{2}(p_{x}^{2} - p_{y}^{2}) + \frac{1}{2}b(x^2 - y^2)
+a xy.
\end{equation}
See that, since the mixed second partial derivative of the potential function does not vanish unless $a\neq 0$, the Hamilton's equation generated by $H$ in \eqref{Hammam} determines a curl force system. 
Further, a direct calculation proves that the Hamilton's equation \eqref{HamEq} reads precisely \eqref{Kapitsa}.
A saddle potential is associated with the equation appears in 
the Euler-Lagrange formulation also. 
After an (inverse) Legendre transformation, it is straight forward to check that the  Lagrangian 
\begin{equation}
 L = \frac{1}{2}(\dot{x}^{2} - \dot{y}^{2}) - \frac{1}{2}b(x^2 - y^2) - axy 
 \end{equation}
yields the model in (\ref{Kapitsa}) as well. A symmetric saddle surface can be described  by ${\tilde U} = (b/2)(x^2 - y^2)$, where $b$
is a geometrical parameter that specifies the curvature of the saddle. An ion trap potential is a rotating
saddle surface on which a ball can be trapped. The ponderomotive potential of an ion trap is that of a saddle in $2D$ but the time evolution 
of potential is one that flaps up and down. The potential of the system is
\begin{equation}
U(x,y,t) = \frac{1}{2}(x^2 - y^2)\cos(2\omega t) -xy \sin(2\omega t),
 \end{equation}
where $\omega$ is the angular drive frequency spinning saddle. 
The system acquires Coriolis and centrifugal forces when written in 
rotating frame and exhibits many interesting mechanics and geometry \cite{KL1,KL2}. 

\section{Curl Forces Coupled with Double Bracket Dissipation}\label{met-Ham}

\subsection{Double Bracket Dissipation}

Linear algebraic dual $\mathfrak{K}^{\ast }$ of a Lie algebra admits (Lie-)Poisson structure, \cite{HoScSt09,demethods2011,LiMa12,MaRa13}. 
For two functions $F$ and $H$, the Lie-Poisson bracket is defined to be 
\begin{equation} \label{LP-Bra}
\{F,H\} ( \textbf{z} ) = \Big\langle \textbf{z} ,\left[ \frac{\delta F}{\delta \textbf{z}},\frac{\delta H}{\delta \textbf{z} }\right]\Big\rangle
\end{equation} 
where $\delta F /\delta \textbf{z} $ is the gradient of  $F$. Here, the bracket on the right hand sider is the Lie algebra bracket on $\mathfrak{K}$. Note that, we assume the reflexivity condition $\mathfrak{K}^{**}=\mathfrak{K}$. 
The dynamics is then computed to be
\begin{equation}\label{aaa}
\dot{F}=\{F, H\} ( \textbf{z}  )
=   \Big\langle \textbf{z} ,\left[ \frac{\delta F}{\delta \textbf{z} },\frac{\delta  H}{
\delta \textbf{z}}\right] \Big\rangle =  \Big\langle \textbf{z} ,- ad_{\delta  H/
\delta \textbf{z}}\frac{\delta F}{\delta \textbf{z} }\Big\rangle =   \Big\langle
 ad_{\delta  H/
\delta \textbf{z}}^{\ast }\textbf{z} ,\frac{\delta F}{\delta \textbf{z} }
\Big\rangle.
\end{equation}
Here, $ ad_\textbf{x} \textbf{x}':=[\textbf{x},\textbf{x}']$ for all $\textbf{x}$ and $\textbf{x}'$ in $\mathfrak{K}$ is the (left) adjoint action of the Lie algebra $\mathfrak{K}$ on itself whereas $ ad^*$ is the (left) coadjoint action of the Lie algebra $\mathfrak{K}$ on the dual space $\mathfrak{K}^*$.
 Notice that $ ad^*_\textbf{x}$ is defined to be minus of the linear algebraic dual of $ ad_\textbf{x}$. Then, 
 we obtain the equation of motion governed by a Hamiltonian function $ H$ as
\begin{equation} \label{eqnofmotion}
\dot{\textbf{z}} -  ad_{{\delta  H}/{\delta \textbf{z} }}^{\ast }\textbf{z}=0.
\end{equation}

\textbf{Coordinate realizations.} Consider a Poisson manifold $\mathcal{P}$ with a local chart $(z_i)$. Then the Poisson bivector $\Lambda=[\Lambda_{ij}]$ determines Poisson bracket as
\begin{equation} \label{PB-local}
\{F, H\}=\Lambda_{ij} \frac{\partial F}{\partial z_i}\frac{\partial  H}{\partial z_i},
\end{equation}
where the summation convention is assumed over the repeated indices. 

Assume a finite dimensional Lie algebra $\mathfrak{K}$ admitting a basis $\{\textbf{k}_{i}\}$ with structure constants $C_{ij}^{k}$ satisfying
\begin{equation}\label{cerceve}
   [\textbf{k}_{i},\textbf{k}_{j}]=C_{ij}^{k}\textbf{k}_{k},
\end{equation}
On the dual basis $\{\textbf{k}^i\}$ of $\mathfrak{K}^*$, we denote an element by $\textbf{z}=z_i\textbf{k}^i$. In this setting, we compute the (Lie-)Poisson bivector as $\Lambda_{ij}=C_{ij}^mz_m$, whereas the Lie-Poisson bracket \eqref{LP-Bra} as
\begin{equation} \label{LP-loc}
\{F, H\}= C^{k}_{ij} z_{k}  \frac {\partial F}{\partial z_{i}} \frac {\partial  H}{\partial z_{j}}.
\end{equation}

\textbf{Metriplectic bracket.} Referring to the (Lie-)Poisson bivector field, define a symmetric bracket, literarily called double bracket, for two functions $F$ and $ H$ as
\begin{equation}\label{doubledissi}
	(F, H)^{(D)}= \mathcal{G}_{il} \frac{\partial F}{\partial z_i} \frac{\partial  H}{\partial z_l}=
\sum_{j}	\Lambda_{ij} \Lambda_{lj}\frac{\partial F}{\partial z_i} \frac{\partial  H}{\partial z_l} =
	\sum_{j}C_{ij}^rC_{lj}^s z_r  z_s \frac{\partial F}{\partial z_i} \frac{\partial  H}{\partial z_l},
\end{equation} 
see \cite{Br93,Mo09}. 
Now we define a metriplectic bracket \cite{Ka84,Ka85,Mo84,Morrison} on $\mathfrak{K}^*$ as the sum of Lie-Poisson bracket and the double bracket \eqref{doubledissi}. Note that, the metriplectic bracket \eqref{db-LP} is an example of a Leibniz bracket
\cite{OrPl04}. In this case, the dynamics of an observable, generated by a Hamiltonian function $H$ and an entropy like function $S$,  is
\begin{equation}\label{db-LP}
\dot{F}=[\vert F, H\vert]^{(D)}=\{F, H\}+a
(F, S)^{(D)},
\end{equation}
for a real number $a$.  Metriplectic systems are particular instances of GENERIC (an acronym for General Equation for Non-Equilibrium Reversible-Irreversible Coupling), \cite{Gr84,GrOt97,PaGr18}.  
We compute the equation of motion as
\begin{equation} \label{aaaa}
\dot{z}_j - C^{m}_{ij} z_{m}\frac{\partial  H}{\partial z_i }= a\sum_{i} C_{ji}^r C_{mi}^n z_r  z_n \frac{\partial  S}{\partial z_m} 
\end{equation}
where on the left hand side we have the reversible Lie-Poisson dynamics whereas the dissipative term is located at the right hand side.

\smallskip

It is not straight forward to derive the dissipative azimuthal curl foces due to mixed
potential term $U(xy)$. We can not use straight away the Herglotz principle and 
the generalized Euler-Lagrange equations to derive dissipative azimuthal curl forces due to
non-separable potential term. We overcome this difficulty by using 
Galley's formalism to study this generalized system and propose an alternative derivation of
dissipative curl force.

\smallskip

Note that the addition of a dissipative
force destabilizes the system regardless of its stability under the action of gyroscopic forces.
This extension overlaps with the program of Krechetnikov and Marsden \cite{KM1,KM}. They proposed a rigorous 
mathematical framework for
the motion of a mechanical system under the influence of various types of forces, namely,
gyroscopic forces, dissipative forces, potential forces and nonconservative positional forces.
We use Galley's formalism to study this generalized system and propose an alternative derivation of
dissipative curl force.

\subsection{Canonical Hamiltonian Dynamics with Double Bracket Dissipation}

Let $V$ be an $n-$dimensional vector space equipped with a basis  $\{\mathbf{e}_i\}$ and its dual $V^*$ with the dual basis $\{\mathbf{e}^i\}$. Then consider a $(2n+1)-$dimensional (Heisenberg) Lie algebra $\mathfrak{g}$ equipped with a basis $\{\mathbf{e}^i,\mathbf{e}_j,\mathbf{f}\}$ where $i$ and $j$ run from $1$ to $n$. Here, the bracket operations are defined to be
\begin{equation}\label{Hei-brack}
[\mathbf{e}^i,\mathbf{e}_j]=\langle \mathbf{e}^i, \mathbf{e}_j\rangle \mathbf{f}= \delta^i_{j}\mathbf{f}
\end{equation} 
and the rest is zero. Here, the second term in  \eqref{Hei-brack} is simply the dualization between $V^*$ and $V$. See that, we can argue the coefficients of this bracket definition as the canonical symplectic form on the symplectic space $V\oplus V^*$.  On the dual space $\mathfrak{g}^*$ we consider the respected dual basis $\{\mathbf{e}_i,\mathbf{e}^j,\mathbf{l}\}$. We denote a Lie algebra element $\boldsymbol{\xi}$ and a dual element $\boldsymbol{\alpha}$  as 
\begin{equation}
\boldsymbol{\xi}=\xi_i\textbf{e}^i+\xi^i\textbf{e}_i+\xi\mathbf{f},\qquad  
\boldsymbol{\mu}=\mu^i\textbf{e}_i+\mu_i\textbf{e}^i+\mu  
\mathbf{l},
\end{equation}
respectively. Obeying the definition presented in \eqref{LP-loc}, the Lie-Poisson bracket of two functions $F$ and $H$ on $\mathfrak{g}^*$   
is computed to be 
\begin{equation}\label{LP-Bra-Hei}
\{F,H\}(\boldsymbol{\mu})=\mu \left( 
\frac{\partial  F}{\partial \mu^i} \frac{\partial  H}{\partial \mu_i}- 
\frac{\partial  H}{\partial \mu^i} \frac{\partial  F}{\partial \mu_i}
\right).
\end{equation}
Referring to this calculation, we write the Lie-Poisson dynamics \eqref{eqnofmotion} generated by a Hamiltonian function $H$ as follows
\begin{equation} \label{Ray-mot}
\dot{\mu}^i-\mu \frac{\partial  H}{\partial \mu_i}=0,  \qquad \dot{\mu}_i+\mu \frac{\partial  H}{\partial \mu^i}=0, \qquad \dot{\mu}=0.
\end{equation} 
Here, the last relation gives that $\mu$ is a constant. In particular, if we choose $\mu^i=q^i$, $\mu_i=p_i$, and $\mu=0$ in the equation (\ref{Ray-mot}) of motion then we arrive at the canonical Hamilton's equations in its very classical form
\begin{equation}
	\dot{q}^i= \frac{\partial  H}{\partial p_i}, \qquad \dot{p}_i=-\frac{\partial H}{\partial q^i}
\end{equation}
but as a coadjoint flow. This naive realization permits us to define a dissipative term to the classical reversible Hamiltonian dynamics by means of a double bracket given in \eqref{doubledissi}. Let us depict this geometry. 

\textbf{Double bracket dissipation.} Referring to the structure constant of the Lie algebra $\mathfrak{g}$  given in \eqref{Hei-brack}, and in the light of the double bracket definition \eqref{doubledissi}, we compute the following symmetric bracket 
\begin{equation}\label{Sym-Bra-Hei}
(F,S)^{(D)}(\boldsymbol{\mu})=\mu   \left(\mu\delta^{ij}\frac{\partial S}{\partial \mu^i}\frac{\partial F}{\partial \mu^j}
+\mu \delta_{ij}
 \frac{\partial S}{\partial \mu_i}\frac{\partial F}{\partial \mu_j}\right ). 
\end{equation}
We add the Lie-Poisson bracket in \eqref{LP-Bra-Hei} and the symmetric bracket in \eqref{Sym-Bra-Hei} and arrive at a metriplectic bracket. The metriplectic dynamics generated by a Hamiltonian function $H$ and an entropy like function $S$ is
 \begin{equation} \label{MD-Ex-1}
 \dot{\mu}^i=\mu \frac{\partial  H}{\partial \mu_i}+\mu \delta^{ij} \mu\frac{\partial S}{\partial \mu^j}, 
 \qquad \dot{\mu}_i=-\mu \frac{\partial  H}{\partial \mu^i}+\mu \mu  \delta_{ij}\frac{\partial \mathcal{S}}{\partial\mu_j},\qquad 
 \dot{\mu}=0.	
 \end{equation}
We wish to record here two interesting particular instances of the metriplectic dynamics \eqref{MD-Ex-1}. For this, once more, we choose $\mu^i=q^i$, $\mu_i=p_i$, and $\mu=1$. 

\textbf{(1)} Let us take the Hamiltonian function $ H=(1/2)\delta^{ij}p_i p_j+V(\textbf{q})$ to be the total energy of the system and $S=S(\textbf{q})$ then the system \eqref{MD-Ex-1} reduces to a second order differential equation with a dissipative term 
  \begin{equation}\label{2ndODE-1}
  	\ddot{q}^i-\frac{\partial^2 S}{\partial q^i  \partial q^j }\dot{q}^j-\delta^{ij}\frac{\partial V}{\partial q^j} =0.
  	 \end{equation}
We cite \cite{Mielke2011} for a more elegant geometrization of the second order ODE \eqref{2ndODE-1}  in terms of the GENERIC framework. Let us now consider that the dimension $n=2$ and consider the Hamiltonian function \eqref{E1a} with $S=-\gamma(t)\delta_{ij}q^iq^j$ then the dynamical system \eqref{MD-Ex-1} reduces to
\begin{equation}
\dot{x}=p_x-\gamma(t)x, \qquad \dot{y}=-p_y-\gamma(t)y, \qquad \dot{p}_x=-xU', \qquad \dot{p}_y=yU'
\end{equation}
where we consider that $\textbf{q}=(q^1,q^2)=(x,y)$. The corresponding Newton's equation is computed to be
\begin{equation}\label{New-1}
\begin{split}
\ddot{x}+\gamma(t)\dot{x}+xU'\big(\frac{1}{2}(x_{1}^{2} - x_{2}^{2}) \big) =0,\\ \ddot{y}+\gamma(t)\dot{y}+yU'\big(\frac{1}{2}(x_{1}^{2} - x_{2}^{2}) \big) =0.
\end{split}
\end{equation}
This pair yields dissipative radial curl forces and determine the Bateman pair of equation. 

\textbf{(2)}  As another application of the dissipative system \eqref{MD-Ex-1}, we consider a Hamiltonian function $H$ and choose $S=(1/2)a\delta^{ij}p_i p_j$ for a scalar $a$, then a fairly straight-forward calculation gives that \eqref{MD-Ex-1} becomes
   \begin{equation}\label{dyn-exp}
\dot{q}^i=\frac{\partial  H}{\partial p_i}, \qquad \dot{p}_i=ap_i-\frac{\partial  H}{\partial q^i}.
  \end{equation}
This is the conformal Hamiltonian dynamics as described in \cite{Perl}. See also \cite{GuhaAGC}. Consider the vector field $X$ generating  the dynamics \eqref{dyn-exp} and observe that, if the canonical symplectic two-form $\Omega=dq^i \wedge dp_i$, 
  \begin{equation}
  \mathfrak{L}_X \Omega= d\big(d H+ap_idq^i\big) = adq^i \wedge dp_i=a \Omega
  \end{equation}	 
  where $\mathfrak{L}$ denotes the Lie derivative.
This syas that $X$ preserves the symplectic two-form up to some conformal factor $a$. Particularly, for the Hamiltonian function \eqref{E1a} in dimension $2$ the dynamical system reduces to conformal curl force system 
\begin{equation}\label{Pre-Bateman}
\dot{x}=p_x, \qquad \dot{y}=-p_y, \qquad \dot{p}_x=-\gamma(t) p_x-xU', \qquad \dot{p}_y=-\gamma(t) p_y+yU'
\end{equation}
where we take $a$ as $-\gamma(t)$. A direct calculation shows that Newtonian realization of this system is exactly the one in \eqref{New-1}.  

\subsection{The Case of Gyroscopic Forces - Magnetic Extension} 

Thomson and Tait classify non-potential forces into three categories \cite{TT}. In $n-$dimensions, consider  the power ${\bf F}\cdot \dot{\bf q}$ of a given system. In the classification, a force $\bf F$ is called gyroscopic force if the power vanishes identically. 
A force field is said to be dissipative if the power is non-positive whereas it is accelerating if the power is non-negative. 

If, further, gyroscopic force is linear then it admits a skew-symmetric coefficient functions $s_{ij}$ so that the components of the force field becomes $F_i = s_{ij}\dot{q}^j$. On the contrary, a linear dissipative force admits 
a symmetric structure $g_{ij}\dot{q}^j$, where the coefficient functions $g_{ij}$ are symmetric. In this case, one can derive the force field by means of a
Rayleigh dissipative function
$R=(1/2)g_{ij}\dot{q}^i\dot{q}^j$. If a Rayleigh function exists the force field is computed to be the gradient of the Rayleigh function with respect to the velocity variables that is $\nabla_{\dot{q}}R$.  

\textbf{Coupling with a Gyroscopic Term.} It is possible to capture gyroscopic forces into a Hamiltonian formalism in a proper Poisson geometry \cite{MaRa13}. See also \cite{Bloch}. To have this, one needs to introduce a non-standard Poisson bracket
 \begin{equation}\label{gyro-bra-}
\{F,H\}_{gyro}(q,p) = \frac{\partial F}{\partial q^i}\frac{\partial H}{\partial p_i} - \frac{\partial F}{\partial p_i} \frac{\partial H}{\partial q^i} - s_{ij} \frac{\partial F}{\partial  p_i}\frac{\partial H}{\partial  p_j},
 \end{equation}
 where the gyroscopic skew-symmetric term involving $\gamma_{ij}$ appears in the ``magnetic extension'' of the bracket. In this case, the dynamics governed by a Hamiltonian function reads
 \begin{equation}\label{gyro-bra}
 \dot{q}^i=\frac{\partial H}{\partial p_i}, \qquad 
 \dot{p}_i=-\frac{\partial H}{\partial q^i}-s_{ij}\frac{\partial H}{\partial  p_j}.
 \end{equation}
 For a Hamiltonian curl force dynamics, similar to the one presented in \eqref{Ham-F}, one can arrive at force field whose coefficient functions are  computed to be linear curl force
$F_i = - \Omega_{ij}q^j$ for a skew-symmetric matrix $\Omega_{ij} = - \Omega_{ji}$. All though, due to skew-symmetric matrix, this looks similar to gyroscopic force theory, it is not the same. In this case, instead of velocities, the force depends on the positions. These forces, particularly, 
are called pseudo gyroscopic of radial corrections coined by Ziegler in 1953. 

It is interesting to examine the Hamiltonian function  \eqref{Ham-F} in the realm of the Poisson bracket \eqref{gyro-bra}. For this, we once more refer to the local coordinates $(x,y)$ on $2-$dimensional base manifold $\mathcal{Q}$. In this case, we take the magnetic term $s_{12}=s$ in the Poisson bracket \eqref{gyro-bra-}. Then the Hamilton's equation \eqref{gyro-bra} generated by the Hamiltonian function \eqref{E1a} turns out to be
\begin{equation}
\dot{x}=p_x,\qquad \dot{y}=-p_y,\qquad \dot{p}_x=-xU'\big(\frac{1}{2}(x^2-y^2) \big)+s p_y, \qquad \dot{p}_y=yU'\big(\frac{1}{2}(x^2-y^2)-s p_x.
\end{equation} 
By collecting all these first order equation in the form of Newton's equation, we conclude that
\begin{equation}\label{Newton-gyro}
\begin{split}
&\ddot{x}+s \dot{y}+xU'\big(\frac{1}{2}(x^2-y^2) \big)=0,\\
 &\ddot{y}-s \dot{x}+yU'\big(\frac{1}{2}(x^2-y^2)=0.
\end{split}
\end{equation}
In this case, the dissipative terms in the system \eqref{Newton-gyro} is given by a skew-symmetric form as a manifestation of the magnetic extension of the Poisson bracket. We can conclude that Hamiltonian of a curl force remains Hamiltonian if the
linear gyroscopic force is added.  

\textbf{Coupling with a Dissipative Term.} The system \eqref{Newton-gyro} includes a curl force and a gyroscopic term. As a further step, we now add a dissipative term to this system. As manifested in the previous section, one can achieve this by employing metriplectic bracket. So we recall the double bracket in \eqref{Sym-Bra-Hei} on a coadjoint orbit and consider only the terms involving partial derivatives with respect to momenta. This time we couple the symmetric bracket with the Poisson bracket \eqref{gyro-bra-} involving a magnetic term. Accordingly, we introduce the following metriplectic bracket
 \begin{equation}
 \begin{split}
[\vert F,H\vert ]_{mb} &= \{F,H\}_{gyro} + c(F,H) \\ &= \frac{\partial F}{\partial q^i}\frac{\partial H}{\partial p_i} - \frac{\partial F}{\partial p_i} \frac{\partial H}{\partial q^i} - s_{ij} \frac{\partial F}{\partial  p_i}\frac{\partial H}{\partial  p_j}+c_{ij}\frac{\partial F}{\partial  p_i}\frac{\partial H}{\partial  p_j},
\end{split}
  \end{equation}
where $s_{ij}$ is skew-symmetric whereas $c_{ij}$ is symmetric. We refer \cite{EsOzSu20} for various couplings of Poisson and symmetric brackets. 

For two dimensional curl force system, we recall the Hamiltonian function in \eqref{Hammam}. We denote the coefficients of the skew-symmetric quantity by $s_{12}=s$ and the coefficients of the symmetric quantity by $c_{11}=c_{22}=c$ while the rest is zero. Then metriplectic equation
\begin{equation}
\dot{\textbf{z}}=[\vert \textbf{z},H\vert ]_{mb} = \{\textbf{z},H\}_{gyro} + c(\textbf{z},H).
\end{equation}
After adding dissipative forces, by means of a symmetric bracket, to the setting, we get the most general
set of system of equations. 
Explicitly, we compute the metriplectic equations as
\begin{equation}
\dot{x}=p_x, \quad \dot{y}=-p_y,\quad \dot{p}_x=-bx-ay+sp_y+cp_x,
\quad \dot{p}_y=by-ax+sp_x-cp_y
\end{equation}
whereas, in the form of Newton's equations, we have that 
\begin{equation}
\ddot{x}+s\dot{y}-c\dot{x}+bx+ay=0, \qquad 
\ddot{y}+s\dot{x}+c\dot{y}+by-ax=0. 
\end{equation}
This pair of equations involve the motion of  mechanical system under influences of various forces which is closely connected
to the work of Krechetnikov and Marsden \cite{KM}, where they
define and study a notion of dissipation  instability in a system ODEs.
Krechetnikov and Marsden \cite{KM} proposed a rigorous mathematical framework for
the motion of a mechanical system under the influence of various types of forces, namely,
gyroscopic forces, dissipative forces, potential forces and nonconservative positional forces. 
Two famous physical examples follow from  reductions of Lagrangian top ($b=0$)
 \begin{equation}
\ddot{x}+s\dot{y}-c\dot{x}+ay=0, \qquad 
\ddot{y}+s\dot{x}+c\dot{y}-ax=0. 
 \end{equation}
Another case is the one where $s=0$. This corresponds dynamics without gyroscopic force term. For this case, the particular instance $a=b=1$ can be written as Euler-Lagrange equations by means of the following Lagrangian function
 \begin{equation}
L = \dot{x}\dot{y} - xy + \frac{1}{2}(x^2 - y^2) + \frac{1}{2}c(\dot{x}y - x \dot{y}).
 \end{equation}
The case $s=0$ and $c=0$ manifests dynamics with no gyroscopic or dissipative force. So this case reduces to the Kapitsa model in \eqref{Kapitsa}.
One must note that the addition of non-conservative positional forces (or curl forces)
to a system with a stable potential energy makes it unstable. The origin of this force
lies in the friction between the rotating shaft and hydrodynamic media.

In 1879 Thomson and Tait \cite{TT} showed that a statically unstable conservative system which has been stabilized by
gyroscopic forces could be destabilized again by the introduction of small damping forces. 
More generally, they consider
conservative and nonconservative linear forces. Note that the instability pops up for a tiny 
bit of damping is
added to the structure. This destabilization paradox is related to the Whitney umbrella 
singularity. See  \cite{Givental,Langford,Nishi,Shapiro} for some works on Whitney umbrella.

\section{Radial Curl Forces in Contact Geometry} \label{Contact}

The Generalized Variational Principle, proposed by Herglotz in 1930, generalizes the classical
variational principle by defining the functional, whose extrema are sought, by
a differential equation. The generalized variational principle gives a variational description of
nonconservative processes. This method provides a link between the mathematical structure of control 
and optimal control theories and contact transformations.
The contact transformations, which can always be derived from the generalized variational principle, 
have found applications in thermodynamics. We use Herglotz principle and the generalized Euler-Lagrange equations
to derive dissipative radial curl forces.

\subsection{Contact Hamiltonian Dynamics}

Let $\mathcal{M}$ be a $(2n+1)-$dimensional manifold with a contact one-form $\sigma \in \Lambda ^{1}\left( \mathcal{M}\right) $ satisfying $d\sigma^n
\wedge \sigma \neq 0$. A contact form determines a contact structure which,
locally is the kernel of the contact form $\sigma$, \cite{arn89,BrCrTa17,ms98,LeLa19}. There is a distinguished (Reeb) vector field satisfying 
\begin{equation}
\iota_{R_{\sigma
}}\sigma =1,\qquad \iota_{R_{\sigma }}d\sigma =0.
\end{equation}
For a Hamiltonian function $H$, the Hamiltonian vector field $X_H$ is the one defined to be 
\begin{equation}\label{Ham-v-f-cont}
\iota_{X_{H}}\sigma =-H, \qquad \iota_{X_{H}}d\sigma =dH-\left(
\iota_{R_{\sigma }}dH\right) \sigma.  
\end{equation}
Being a non-vanishing top-form we can consider $d\sigma^n
\wedge \sigma $ as a volume form on $\mathcal{M}$. It is important to note that the Hamiltonian motion does not preserve the volume form. 
In this realization, contact Poisson (or Lagrange) bracket of two smooth functions on $\mathcal{M}
$ is defined by
\begin{equation}\label{cont-bracket}
\{F,H\}=\iota_{[X_F,X_H]}\sigma,
\end{equation}
 where $X_F$ and $X_H$ are Hamiltonian vectors fields determined through \eqref{Ham-v-f-cont}. 

\textbf{Darboux' Coordinates.} There are Darboux' coordinates on $\mathcal{M}$ given by $(q^i,p_i,z)$. In this realization, the contact one-form is $\sigma =dz-p_idq^i$ and the Reeb vector field is  $R_{\sigma }=\partial /\partial z$. For a Hamiltonian function $H$, the Hamiltonian vector field, determined in \eqref{Ham-v-f-cont}, is computed to be
\begin{equation}\label{con-dyn}
X_H=\frac{\partial H}{\partial p_i}\frac{\partial}{\partial q^i}  - \big (\frac{\partial H}{\partial q^i} + \frac{\partial H}{\partial z} p_i \big)
\frac{\partial}{\partial p_i} + (p_i\frac{\partial H}{\partial p_i} - H)\frac{\partial}{\partial z},
\end{equation}
whereas the contact Poisson bracket \eqref{cont-bracket} is 
\begin{equation}
\{F,H\} = \frac{\partial F}{\partial q^i}\frac{\partial H}{\partial p_i} -
\frac{\partial F}{\partial p_i}\frac{\partial H}{\partial q^i} + \big(F  - p_i\frac{\partial F}{\partial p_i} \big)\frac{\partial H}{\partial z} -
\big(H  - p_i\frac{\partial H}{\partial p_i} \big)\frac{\partial F}{\partial z}.
\end{equation}
We obtain the contact Hamiltonian system
\be \label{conham}
\dot{q}^i= \frac{\partial H}{\partial p_i}, \qquad \dot{p}_i = -\frac{\partial H}{\partial q^i}- 
p_i\frac{\partial H}{\partial z}, \quad \dot{z} = p_i\frac{\partial H}{\partial p_i} - H.
 \ee

\subsection{Herglotz principle}

The generalized variational principle, proposed by Herglotz, defines the functional whose
extrema are obtained by a differential equation rather than by an integral.
The Herglotz principle yields a variational description of nonconservative as well as conservative
processes involving one independent variable.
The Herglotz principle is defined by the functional $z({\bf q};\tau)$
through a differential equation of the form \cite{Guether,Herglotz,LeLa21}
\be\label{herglotzprinciple} \dot{z} = L(t,q^i,\dot{q}^i,z), \qquad 0 \leq t \leq \tau. \ee
Here $\bf q$ belongs to the space of $C^2$ curves $\bf q$ defined on $[0,\tau]$
satisfying the boundary condition ${\bf q}(0) = {\bf q}_0$, ${\bf q}(\tau) = {\bf q}_{\tau}$. So $z$ is a solution of the Cauchy problem
of Herglotz action with $z(0)= z_0$.

\bigskip

Let $L$ be a Lagrangian function defined on $T\mathcal{Q}\times \mathbb{R}$ with coordintes $(q^i,\dot{q}^i,z)$. Herglotz showed that the value of this
functional attains its extremum if  ${\bf q}(t)$ solution of the generalized
Euler-Lagrange equations (the dissipative Lagrange system)
\be\label{Herglotz}
\frac{\partial L}{\partial q^i} - \frac{d}{dt}\Big(\frac{\partial L}{\partial {\dot q}^i} \Big)
+ \frac{\partial L}{\partial z}\frac{\partial L}{\partial {\dot q}^i} = 0, 
\ee
for all $t \in [0, \tau]$ and $z$ is a solution of the Cauchy problem (\ref{herglotzprinciple})
which depends on $\bf{q}$. 
It is important to notice that (\ref{Herglotz}) represents a family of 
differential equations since for each
function $\bf{q}(t)$ a different differential equation arises, hence $z(t)$ dependes on ${ \bf q}(t)$.
Without the dependence of $z$, this problem reduces to a classical calculus of variations problem. If the functional $z$ defined in \eqref{herglotzprinciple} is invariant 
with respect to translation in time, then the quantity
\be \label{consrule}
I = exp\Big(- \int^t  \frac{\partial L}{\partial z} d\theta \Big)\Big(L ( x, \dot{x}, z) - 
\frac{\partial L}{\partial \dot{x}^k}\dot{x}^k \Big)
\ee
is conserved on solutions of the generalized Euler-Lagrange equations for regular Lagrangians.

\textbf{The Legendre Transformation.} For a regular Lagrangian function $L$ we define the fiber derivative is 
\begin{equation}\label{Leg-Trf}
\mathbb{F}L: T\mathcal{Q}\times \mathbb{R}\longrightarrow  T^*\mathcal{Q}\times \mathbb{R}, \qquad (q^i,\dot{q}^i,z)\mapsto 
(q^i,\frac{\partial L}{\partial \dot{q}^j},z)
\end{equation}
A direct calculation shows that the Legendre transformation \eqref{Leg-Trf} maps the generalized Euler-Lagrange equations in \eqref{Herglotz} to contact Hamiltonian dynamics \eqref{conham}.

\smallskip

The flow of a contact Hamiltonian system preserves the contact structure, but it does not preserve the Hamiltonian. Instead we obtain
$$
\frac{dH}{dt} = - H\frac{\partial H}{\partial z}. $$
It can be readily checked, in the light of the integral invariant \eqref{consrule}, that
\be I(x, p, z) = H(x(t), p(t), z(t)) exp \Big(\int^t \frac{\partial H}{\partial z} d\theta \Big) \ee
is constant along the flow of $X_{H}^{c}$  defined by (\ref{conham}) with the autonomous contact Hamiltonian
$H(x, p, z)$.

\subsection{Radial Curl Forces in Contact Dynamics}

Consider two-dimensional anisotropic damped system with the potential energy $ \frac{1}{2}(x_{1}^{2} - x_{2}^{2})$ and changing-sign 
damping coefficient. The contact Hamiltonian is recasted as 
 \begin{equation}\label{cont-Ham-Func}
H = \frac{1}{2} (p_{x}^{2} - p_{y}^{2} ) + U\big(\frac{1}{2}(x^{2} - y^{2}) \big)
+ \gamma(t) z
 \end{equation}
 and leads to the equations
 \begin{equation} \label{eq-cont}
 \begin{split}
 \dot{x} &= p_x, \quad \dot{y} = -p_y, \\ \dot{p} _x&= -x_1 
U^{\prime}\big(\frac{1}{2}(x_{1}^{2} - x_{2}^{2}) \big) -\gamma(t) p_x, , \\ \dot{p}_y &= x_2 
U^{\prime}\big(\frac{1}{2}(x_{1}^{2} - x_{2}^{2}) \big) + \gamma(t)p_y, \\
\dot{z}&=\frac{1}{2} (p_{x}^{2} - p_{y}^{2} ) - U\big(\frac{1}{2}(x^{2} - y^{2}) \big)
- \gamma(t) z.
 \end{split}
 \end{equation}
Notice that the first two lines of equations are precisely the dissipative dynamics in \eqref{Pre-Bateman}. So that, they are equal to the Bateman pair of equations in \eqref{New-1}. In order to investigate the last equation in \eqref{eq-cont}, we first recall  the inverse Legendre transformation
\begin{equation}
\mathbb{F}H:T^*\mathcal{Q}\times \mathbb{R} \longrightarrow 
T\mathcal{Q}\times \mathbb{R}, (q^i,p_i,z)\mapsto (q^i,\frac{\partial H}{\partial p_i},z).
\end{equation}
Evidently, the Hamiltonian function \eqref{cont-Ham-Func} has indeed a regular (invertible) Legendre transformation. The first set of equations in \eqref{eq-cont}  is the realization of the inverse Legendre trasnformation. Further, it is immediate to see that by imposing the last equation in \eqref{eq-cont} as a differential Herglotz principle for the Lagrangian function 
\begin{equation}
L(x,y,\dot{x},\dot{y})=\frac{1}{2} (\dot{x}^{2} - \dot{y}^{2} ) - U\big(\frac{1}{2}(x^{2} - y^{2}) \big)- \gamma(t) z.
\end{equation}
Then, in the light of the inverse Legendre transformation $\dot{x}=p_x$ and $\dot{y} = -p_y$, the first two lines of the system \eqref{eq-cont} are the generalized (dissipative) Euler-Lagrange equations in \eqref{Herglotz}. This observation manifests the variational aspect of the Bateman's pair \eqref{New-1} due the radial curl forces with a dissipation.   

\textbf{Kapitsa-Merkin Model with Dissipation.} Recall the Hamiltonian realization \eqref{Hammam} introduced for the Kapitsa model in \eqref{Kapitsa}. To generalize this discussion to the contact Hamiltonian framework, introduce the following 
contact Hamiltonian function 
\begin{equation}
H = \frac{1}{2}(p_{x}^{2} - p_{y}^{2}) + \frac{1}{2}b(x^2 - y^2)
+a xy+ \gamma(t)z.
 \end{equation}
A direct computation gives that the first two sets of equations in the contact Hamilton's equations \eqref{conham} can be written in  the form of Newton's equation
\begin{equation}
\ddot{x}+\gamma(t)\dot{x}+bx+ay=0,\qquad \ddot{y}+\gamma(t)\dot{y}+by-ax=0.
 \end{equation}
The last equation in \eqref{conham} becomes the differential equation
\begin{equation}
\dot{z}=\frac{1}{2}(\dot{x}^{2} - \dot{y}^{2}) - \frac{1}{2}b(x^2 - y^2)
- a xy- \gamma(t)z
 \end{equation}
determining the Herglotz principle \eqref{herglotzprinciple}.

\section{Dissipative Curl Forces and Galley's Method}\label{Galley-Sec}

Hamilton’s action principle is not suitable for non-conservative systems. Having observed that Galley \cite{Galley,Galley1} proposed a method based on a new variational principle. 
This required allowing one to break the time-symmetry manifest in the action.
The formalism given in \cite{Galley}
corresponds to a variational principle specified by initial
data contrary to Hamilton’s that fixes
configuration of system at initial and final
times. We first depict this geometry then employ this approach to curl force theory.

\subsection{Galley's Formalism}

Galley has developed a consistent formulation of Hamilton’s principle that is compatible
with initial value problems \cite{Galley,Galley1}. In the framework of Galley, the degrees
of freedom are formally doubled to facilitate the
nonconservativity, so that ${\bf q}_1$ and ${\bf q}_2$ are decoupled from each other. Accordingly, one defines an action, a functional of the coordinates ${\bf q}_1$ and ${\bf q}_2$, as
\begin{equation}
S({\bf q}_1,{\bf q}_2) = \int_{t_i}^{t_f}dt{\cal L}({\bf q}_1, {\bf q}_2, {\bf \dot{q}}_1, {\bf \dot{q}}_2),
\end{equation}
where the Lagrangian is determined in the form of
\begin{equation}
{\cal L}({\bf q}_1, {\bf q}_2, {\bf \dot{q}}_1, {\bf \dot{q}}_2) = L({\bf q}_1, {\bf \dot{q}}_1) - L({\bf q}_2, {\bf \dot{q}}_2) + 
K({\bf q}_1, {\bf q}_2, {\bf \dot{q}}_1, {\bf \dot{q}}_2),
\end{equation}
where $K({\bf q}_1, {\bf q}_2, {\bf \dot{q}}_1, {\bf \dot{q}}_2)$ encodes nonconservative nature of a dynamical system. It is immediate to see that $K$ must be skew-symmetric with respect to the relabelling ${\bf q}_1\leftrightarrow {\bf q}_2$. 
For $K \neq 0$ the two paths ${\bf q}_1(t)$ and ${\bf q}_2(t)$ get coupled with each other and ${\cal L}$
describes a non conservative system. The Euler-Lagrange equations for  ${\cal L}$ are not necessarily physical until we take the physical limit (PL)
wherein the histories are identified, i.e.,
${\bf q}_1 = {\bf q}_2 = {\bf q}$, after completing all variations and derivatives.
A more convenient parametrization of the coordinates is exhibited in \cite{Galley} as
\begin{equation}\label{new-coord}
{\bf q}_- = 
{\bf q}_1 - {\bf q}_2, \qquad  {\bf q}_+ = \frac{{\bf q}_1 + {\bf q}_2}{2}.
\end{equation}
It is evident that, after taking the physical limit PL, one has that  \begin{equation}\label{PL}
{\bf q}_- \to 0, \qquad {\bf q}_+ \to {\bf q}_1= {\bf q}_2=\bf q .
\end{equation}
This leads to the dissipative Euler–Lagrange equations
\begin{equation} \label{GalleyEL}
\frac{d}{dt}\left(\frac{\partial L}{\partial \bf \dot{q}} + \Big[\frac{\partial K}{\partial {\bf \dot{q}_-}}\Big]_{PL} \right )= \frac{\partial L}{\partial {\bf q}} +
\Big[\frac{\partial K}{\partial {\bf q}_-}\Big]_{PL}. 
\end{equation}

Consider a regular Lagrangian function $L=L({\bf q}_a,\dot{\bf q}_a)$ for $a=1,2$. In terms of the Legendre transformation ${\bf p}_a=\partial L / \partial {\bf \dot{q}}_a$, we define the canonical Hamiltonian function  $H({\bf q}_a,{\bf p}_a)={\bf p}_a\cdot{\bf \dot{q}}_a-L$. This reads the following collective Hamiltonian function 
\begin{equation}
{\cal H} = H({\bf q}_1,{\bf p}_1) -  H({\bf q}_2,{\bf p}_2) - K({\bf q}_1, {\bf q}_2, {\bf p}_1 ,{\bf p}_2).
\end{equation}
It is possible to write the Hamiltonian function in terms of the variables in \eqref{new-coord}. So that we can compute the momenta $\bf{p}_{+}$ and $\bf{p}_{-}$ so that we can write the Hamiltonian function with respect to $\pm$ coordinates that is ${\cal H}={\cal H}({\bf q}_-,{\bf q}_+,\bf{p}_{-},\bf{p}_{+})$. After taking the physical limit PL in \eqref{PL}, the canonical Hamilton's equations reduce to
\begin{equation}\label{Galleyeqn} 
{\dot{\bf q}} = \frac{\partial H}{\partial {\bf p}} - \Big[\frac{\partial K}{\partial {\bf p}_-}\Big]_{PL}, \qquad 
{\dot{\bf p}} = -\frac{\partial H}{\partial {\bf q}} + \Big[\frac{\partial K}{\partial {\bf q}_-}\Big]_{PL}.
\end{equation}
Non-conservative character of this system can be easily observed by the time derivative of the Hamiltonian function 
\begin{equation}
\frac{dH}{dt} = - {\bf q}\cdot \Big[\frac{d}{dt}\frac{\partial K}{\partial \dot{\bf q}_-} - \frac{\partial K}{\partial {\bf q}_-}\Big]_{PL}.
\end{equation}

\subsection{Dissipative Curl Forces in Galley's Framework}

In this section, we employ the dissipative decoupling method presented in the previous subsection to the Hamiltonian functions governing radial curl forces. 

\textbf{Decoupling Bateman's Pair.} 
We start with the Hamiltonian function given in \eqref{E1a} where the potential energy is particularly considered to be $U=(1/2)(x^2 - y^2)$. Decompose this Hamiltonian function as
 \begin{equation} \label{lincurlEq} 
H = \frac{1}{2}(p_{x}^{2} - p_{y}^{2}) + \frac{1}{2}(x^2 - y^2) = H_1(x,p_x) - H_2(y,p_y), 
 \end{equation}
where $H_1 =(1/2) p_{x}^{2} +(1/2) x^2$. We now couple the Hamiltonian function $H$ in \eqref{lincurlEq} with a function $K=\kappa p_+x_-$ where the plus minus notation is adopted through \eqref{new-coord} so that $x_+=(1/2)(x+y)$ and $x_-=x-y$. Explicitly, we have that
  \begin{equation}
  {\cal H} (x_-,x_+,p_-,p_+)= H(x_-,x_+,p_-,p_+) - K(x_-,x_+,p_-,p_+)=p_-p_++x_-,x_+-\kappa p_+x_-,
  \end{equation} 
where $H$ is the one in \eqref{lincurlEq} with the plus minus coordinates. The non-conservative Hamilton's equation in \eqref{Galleyeqn} read that
\begin{equation}
\dot{x} = p_x, \qquad \dot{p}_x = -x + \kappa p_x
\end{equation}
whereas the Newtonian form of the dynamics is 
\begin{equation}
\ddot{x} - \kappa \dot{x} + x = 0.
\end{equation}
This observation explores that variational underlying the individual dynamics in the Bateman's pair is computed through the coupled Lagrangian 
\begin{equation}
L(x,y,\dot{x},\dot{y})=\frac{1}{2}(\dot{x}^{2} - \dot{y}^{2}) - \frac{1}{2}(x^2 - y^2)-\frac{1}{2}(\dot{x}x-\dot{x}y+\dot{y}x-\dot{y}y).
\end{equation} 
according to the nonconservative Euler-Lagrange equations \eqref{GalleyEL}. 

\textbf{Dissipation added to Curl Forces. }
In this case we apply Galley's strategy to couple dissipation to curl forces. Start with the following Hamiltonian function, on the cotangent bundle of two-dimensional Euclidean space, 
\begin{equation}
H  =  \frac{1}{2}(p_{x}^{2} - p_{y}^{2}) + \frac{1}{2}b(x^2 - y^2) +a xy ,
\end{equation} 
where $a$ and $b$ are real constants. 
Then, we couple the coordinates $(x,y,p_x,p_y)$ with $(u,v,p_u,p_v)$, and define the plus minus notation
\begin{equation}
\begin{split}
\textbf{q}_-&=(x-u,y-v),\qquad \textbf{q}_+=\big(\frac{1}{2}
(x+u),\frac{1}{2}(y+v)\big), \\ \textbf{p}_-&=(p_x-p_u,p_y-p_v), 
\qquad 
\textbf{p}_+=\big(\frac{1}{2}
(p_x+p_u),\frac{1}{2}(p_y+p_v)\big). 
\end{split}
\end{equation}
In order to add a dissipation to the Hamiltonian dynamics, we introduce 
\begin{equation}
K = -\kappa {\bf p}_+ \cdot {\bf q }_- + \textbf{f}(t)\cdot {\bf q}_-,
\end{equation}
where $\textbf{f}(t)$ stands for an external force. 
Evaluating at the physical limit, the dissipative Hamilton's equation in \eqref{Galleyeqn} reads the equation of motion for a forced damped
linear curl system in the Hamiltonian form
\begin{equation}
\dot{x}=p_x,\qquad \dot{y}=-p_y,\qquad \dot{p}_x=-bx-ay-\kappa p_x+ f_x(t), \qquad \dot{p}_y=by-ax-\kappa p_y+f_y(t).
\end{equation}
whereas in the Newtonian form one arrives at
\begin{equation}
\ddot{x} + \kappa \dot{x} + bx + ay = f_x(t), \qquad \ddot{y} + \kappa \dot{y} + by -ax = f_y(t).
\end{equation}
The same result will be exhibited from Galley's equation by adding, for example a term if form
$\textbf{p}_+ \cdot \textbf{q}_+$ term since such terms do not contribute anything
to the Galley's  equation \eqref{Galleyeqn}. But this form is clearly connected to the Bateman term appears in 
the class of quadratic Hamiltonians of damped curl forces
Observe that, if the the external force is zero and $\kappa$ vanishes then one arrives precisely at the Kapitsa model in \eqref{Kapitsa}.

\section{Conclusion}
We have explored several sets of ideas centered around linear Hamiltonian curl forces as proposed by Berry and Shukla.
We have used double bracket dissipation method, contact Hamiltonian formalism and Galley's method to derive the dissipative extension of the curl force. It is noteworthy that the simplecticity is also destroyed by 
an addition of non-conservative positional forces. Double bracket formalism is manifested in a metriplectic dynamics. The addition of nonzero dissipative curl force can be described via metriplectic structure. For the linear case, we have investigated the steps starting from linear curl force through the 
Krechetnikov-Marsden \cite{KM} type equation by incorporating dissipation and gyroscopic forces.

\section*{Acknowledgements}
We would like to express our sincere appreciation to
Professors Sir Michael Berry, Anindya Ghose-Choudhury, Jayanta Bhattacharjee, 
Pavle Saksida for their interest, encouragement and valuable comments in various stages of our work.
This work was done while (PG)  was visiting Department of Mathematics, Gebze Technical University
under T\"ubitak visiting professorship.
(PG) would like to express his gratitude to the members of the department for their warm hospitality,


\begin{thebibliography}{99}

\bibitem{arn89} 
Arnol'd, V. I. (2013). Mathematical methods of classical mechanics (Vol. 60). Graduate Texts in Mathematics 60,
(Springer-Verlag, 1989).



\bibitem{AML} Albaladejo, S., Marqués, M. I., Laroche, M., \& Sáenz, J. J. (2009). Scattering forces from the curl of the spin angular momentum of a light field. Physical review letters, 102(11), 113602.

\bibitem {deAzPeBu96}
de Azcárraga, J. A., Perelomov, A. M., \& Bueno, J. P. (1996). The Schouten-Nijenhuis bracket, cohomology and generalized Poisson structures. Journal of Physics A: Mathematical and General, 29(24), 7993.

\bibitem{BS466} Berry, M. V., \& Shukla, P. (2012). Classical dynamics with curl forces, and motion driven by time-dependent flux. Journal of Physics A: Mathematical and Theoretical, 45(30), 305201.


\bibitem{BS3} Berry, M. V., \& Shukla, P. (2015). Hamiltonian curl forces. Proceedings of the Royal Society A: Mathematical, Physical and Engineering Sciences, 471(2176), 20150002.

\bibitem{BS2} Berry, M. V., \& Shukla, P. (2016). Curl force dynamics: symmetries, chaos and constants of motion. New Journal of Physics, 18(6), 063018.


\bibitem{Bloch} Bloch, A. M., Hagerty, P., Rojo, A. G., \& Weinstein, M. I. (2004). Gyroscopically stabilized oscillators and heat baths. Journal of statistical physics, 115(3), 1073-1100.

\bibitem{BrCrTa17}
Bravetti, A., Cruz, H., \& Tapias, D. (2017). Contact hamiltonian mechanics. Annals of Physics, 376, 17-39.


\bibitem {Br93} 
Brockett, R. W. (1993, March). Differential geometry and the design of gradient algorithms. In Proc. Symp. Pure Math., AMS (Vol. 54, No. 1, pp. 69-92).


\bibitem{CNV} Chaumet, P. C., \& Nieto-Vesperinas, M. (2000). Time-averaged total force on a dipolar sphere in an electromagnetic field. Optics letters, 25(15), 1065-1067.

\bibitem{Devaney}  Devaney, R. L. (1978). Nonregularizability of the anisotropic Kepler problem. Journal of Differential Equations, 29(2), 253-268.

\bibitem{EsOzSu20}
Esen, O., Özcan, G., \& Sütlü, S. (2021). On Extensions, Lie-Poisson Systems, and Dissipations. arXiv preprint arXiv:2101.03951.


\bibitem{Galley} Galley, C. R. (2013). Classical mechanics of nonconservative systems. Physical review letters, 110(17), 174301.


\bibitem{Galley1} Galley  C R, Tsang D \& Stein L C, 2014, The principle of stationary nonconservative action
for classical mechanics and field theories,  arXiv:1412.3082, 2014.

\bibitem{Givental} Givental’, A. B. (1986). Lagrangian imbeddings of surfaces and unfolded Whitney umbrella. Functional Analysis and its applications, 20, 197-203.

\bibitem{GCGP} Ghose-Choudhury, A., Guha, P., Paliathanasis, A., \& Leach, P. G. L. (2017). Noetherian symmetries of noncentral forces with drag term. International Journal of Geometric Methods in Modern Physics, 14(02), 1750018.

\bibitem{Gr84}
Grmela, M. (1984). Bracket formulation of dissipative fluid mechanics equations. Physics Letters A, 102(8), 355-358.

\bibitem {GrOt97}
 Grmela, M., \& \"Ottinger, H. C. (1997). Dynamics and thermodynamics of complex fluids. I. Development of a general formalism. Physical Review E, 56(6), 6620--6632.
 
 
\bibitem{Guha} Guha, P. (2007). Metriplectic structure, Leibniz dynamics and dissipative systems. Journal of Mathematical Analysis and Applications, 326(1), 121-136.

\bibitem{Guether} Guenther, R. B., Schwerdtfeger, H., Herglotz, G., Guenther, C. M., \& Gottsch, J. A. (1996). The Herglotz lectures on contact transformations and Hamiltonian systems. Juliusz Schauder Center for Nonlinear Studies. Nicholas Copernicus University.

\bibitem{Guha1} Guha, P. (2018). Saddle in linear curl forces, cofactor systems and holomorphic structure. The European Physical Journal Plus, 133(12), 536.

\bibitem{Guha2} Guha, P. (2020). Curl forces and their role in optics and ion trapping. The European Physical Journal D, 74, 1-12.

\bibitem{GuhaAGC} Guha, P., \& Ghose-Choudhury, A. (2018). Generalized conformal Hamiltonian dynamics and the pattern formation equations. Journal of Geometry and Physics, 134, 195-208.

\bibitem{Gumral} Gümral H (2018) Poisson Bi-vectors in Four Dimensions, preprint.

\bibitem{Gutz} Gutzwiller, M. C. (1973). The anisotropic Kepler problem in two dimensions. Journal of Mathematical Physics, 14(1), 139-152.

\bibitem{HoScSt09}
Holm, D. D., Schmah, T., \& Stoica, C. (2009). Geometric mechanics and symmetry: from finite to infinite dimensions.


\bibitem{Herglotz} Herglotz G. (1930). Ber\"uhrungstransformationen, Lectures at the University of G\"ottingen, 
G\"ottingen.

\bibitem{Kapitsa} Kapitsa, P. L. (1939). Stability and transition through the critical speed of fast rotating 
shafts with friction. Zhur. Tekhn. Fiz. 9 124-147.

\bibitem{KapitsaPen} Kapitsa, P. L. (1951) Dynamic stability of a pendulum when its point of suspension vibrates". Soviet Phys. JETP. 21: 588-592.; Kapitza P L, 1951, Pendulum with a vibrating suspension, Usp. Fiz. Nauk. 44: 7-15.

\bibitem {Ka84} 
Kaufman, A. N. (1984). Dissipative Hamiltonian systems: a unifying principle. Physics Letters A, 100(8), 419-422.

\bibitem{Ka85}
 Kaufman, A. N. (1985). Lorentz-covariant dissipative Lagrangian systems. Physics Letters A, 109(3), 87-89.
 
\bibitem{KL2} Kirillov, O. N., \& Levi, M. (2017). Rotating saddle trap: A Coriolis force in an inertial frame, 
Nonlinearity 30, 1109-1119.
 
\bibitem{KL1} Kirillov, O. N., \& Levi, M. (2016). Rotating saddle trap as Foucault's pendulum. American Journal of Physics, 84(1), 26-31.


\bibitem{KM1} Krechetnikov, R., \& Marsden, J. E. (2006). On destabilizing effects of two fundamental non-conservative forces. Physica D: Nonlinear Phenomena, 214(1), 25-32.

\bibitem{KM} Krechetnikov, R., \& Marsden, J. E. (2007). Dissipation-induced instabilities in finite dimensions. Reviews of modern physics, 79(2), 519.

\bibitem{Langford} Langford, W. F. (2003). Hopf meets Hamilton under Whitney’s umbrella. In IUTAM Symposium on Nonlinear Stochastic Dynamics (pp. 157-165). Springer, Dordrecht. 

\bibitem {LaPi12} 
Laurent-Gengoux, C., Pichereau, A., \& Vanhaecke, P. (2012). Poisson structures (Vol. 347). Springer Science \& Business Media.


\bibitem{LeLa19}
de León, M., \& Lainz Valcázar, M. (2019). Contact Hamiltonian systems. Journal of Mathematical Physics, 60(10), 102902.

\bibitem{LeLa21}
de León, M., \& Lainz, M. (2020). A review on contact Hamiltonian and Lagrangian systems. arXiv preprint arXiv:2011.05579.


\bibitem{demethods2011} 
de León, M., \& Rodrigues, P. R. (2011). Methods of differential geometry in analytical mechanics. Elsevier.

\bibitem {LiMa12} 
Libermann, P., \& Marle, C. M. (2012). Symplectic geometry
and analytical mechanics (Vol. 35). Springer Science \& Business Media.




\bibitem{Lundmark2} Lundmark, H. (2003). Higher‐Dimensional Integrable Newton Systems with Quadratic Integrals of Motion. Studies in Applied Mathematics, 110(3), 257-296.

\bibitem{Lundmark3} Lundmark, H. (1999). Integrable nonconservative Newton systems with quadratic integrals of motion. Matamatiska Institutionen, Linköpings Universitet. Linköping Studies in Science and Technology, Thesis No. 756, Link\"oping Univ., Link\"oping.

\bibitem{ms98} McDuff, D., \& Salamon, D. (2017). Introduction to symplectic topology. Oxford University Press.

\bibitem{Perl} 
McLachlan, R., \& Perlmutter, M. (2001). Conformal Hamiltonian systems. Journal of Geometry and Physics, 39(4), 276-300.

\bibitem{MaRa13}
Marsden, J. E., \& Ratiu, T. S. (2013). Introduction to mechanics and symmetry: a basic exposition of classical mechanical systems.




\bibitem{Merkin} Merkin, D. R. (1974). Gyroscopic systems. Moscow Izdatel Nauka.


\bibitem{Merkin1} Merkin, D. R. (2012). Introduction to the Theory of Stability (Vol. 24). Springer Science \& Business Media.

\bibitem{Mielke2011} Mielke, A. (2011). Formulation of thermoelastic dissipative material behavior using GENERIC. Continuum Mechanics and Thermodynamics, 23(3), 233-256.

\bibitem {Mo84} 
Morrison, P. J. (1984). Bracket formulation for irreversible classical fields. Physics Letters A, 100(8), 423-427.


\bibitem{Morrison} Morrison, P. J. (1986). A paradigm for joined Hamiltonian and dissipative systems. Physica D: Nonlinear Phenomena, 18(1-3), 410-419.

\bibitem {Mo09} 
Morrison, P. J. (2009). Thoughts on brackets and dissipation: old and new. In Journal of Physics: Conference Series (Vol. 169, No. 1, p. 012006). IOP Publishing.


\bibitem{Nishi} Nishimura, T. (2011). Whitney umbrellas and swallowtails. Pacific journal of mathematics, 252(2), 459-471.


\bibitem {OrPl04}
 Ortega, J. P., \& Planas-Bielsa, V. (2004). Dynamics on Leibniz manifolds. Journal of Geometry and Physics, 52(1), 1-27.
 
 
\bibitem{Paul} 
Paul, W. (1990). Electromagnetic traps for charged and neutral particles. Reviews of modern physics, 62(3), 531.

\bibitem{PaGr18}
Pavelka, M., Klika, V., \& Grmela, M. (2018). Multiscale thermo-dynamics: introduction to GENERIC. Walter de Gruyter GmbH \& Co KG.

\bibitem{Lundmark1} Rauch-Wojciechowski, S., Marciniak, K., \& Lundmark, H. (1999). Quasi-Lagrangian systems of Newton equations. Journal of Mathematical Physics, 40(12), 6366-6398.



\bibitem{Shapiro} Shapiro, B. Z., \& Khesin, B. A. (1992). Swallowtails and Whitney umbrellas are homeomorphic. J. Alg. Geom, 1, 549-560.

\bibitem{SS}  Shimizu, Y., \& Sasada, H. (1998). Mechanical force in laser cooling and trapping. American Journal of Physics, 66(11), 960-967.


\bibitem{Stephenson} Stephenson A, 1908. XX.On induced stability, Philosophical Magazine. 6. 15: 233-236.
\bibitem{TT} Thomson, S. W., \& Tait, P. G. (1867). Treatise on natural philosophy, Vol. I. Oxford : Clarendon Press.



\bibitem {Va12}
Vaisman, I. (2012). Lectures on the geometry of Poisson manifolds (Vol. 118). Birkhäuser.

\bibitem {wei83}
 Weinstein, A. (1983). The local structure of Poisson manifolds. Journal of differential geometry, 18(3), 523-557.
 
\end{thebibliography}
\end{document}